\documentclass[letterpaper]{article}

\usepackage[T1]{fontenc}

\usepackage{geometry}
\usepackage{setspace}

\usepackage[style=numeric, sorting=none, maxbibnames=99, giveninits=true]{biblatex}
\usepackage{graphicx}
\usepackage{float}
\newfloat{scheme}{htbp}{los}
\floatname{scheme}{Scheme}
\floatname{chart}{Chart}
\newfloat{graph}{htbp}{loh}

\usepackage{chemformula} 
\usepackage[version = 4]{mhchem} 

\usepackage{amssymb}
\usepackage{subcaption} 
\usepackage[colorinlistoftodos]{todonotes}
\usepackage{cleveref}
\usepackage{algorithm}
\usepackage{algpseudocode}
\usepackage{bbm}
\usepackage{booktabs} 
\usepackage{longtable} 

\DeclareFieldFormat{labelnumberwidth}{#1.}

\DeclareFieldFormat[article,inproceedings]{title}{#1}
\DeclareFieldFormat{journaltitle}{\mkbibemph{#1}}
\DeclareFieldFormat{year}{\mkbibbold{#1}}
\DeclareFieldFormat{volume}{\mkbibemph{#1}}
\renewbibmacro*{journal+issuetitle}{%
  \usebibmacro{journal}%
  \setunit*{\addspace}%
  \printfield{year}%
  \setunit*{\addcomma\space}%
  \printfield{volume}%
  \setunit*{\addcomma\space}%
}
\DeclareFieldFormat{pages}{#1}

\usepackage{authblk}
\author[1,4]{Yannik Mahlau*}
\author[4]{Yannick Augenstein}
\author[4]{Tyler W. Hughes}
\author[2,3]{Marius Lindauer}
\author[1,3]{Bodo Rosenhahn}
\affil[1]{Institute of Information Processing, Leibniz University Hannover, Germany}
\affil[2]{Institute of Artificial Intelligence, Leibniz University Hannover, Germany}
\affil[3]{L3S Research Center, Hannover, Germany}
\affil[4]{Flexcompute, Watertown, MA, USA}

\title{Gradient-Informed Bayesian and Interior Point Optimization for Efficient Inverse Design in Nanophotonics}

\date{*Email: mahlau@tnt.uni-hannover.de}

\begin{document}
\maketitle

\begin{abstract}
Inverse design, particularly geometric shape optimization, provides a systematic approach for developing high-performance nanophotonic devices.
While numerous optimization algorithms exist, previous global approaches exhibit slow convergence and conversely local search strategies frequently become trapped in local optima.
To address the limitations inherent to both local and global approaches, we introduce BONNI: Bayesian optimization through neural network ensemble surrogates with interior point optimization.
It augments global optimization with an efficient incorporation of gradient information to determine optimal sampling points.
This capability allows BONNI to circumvent the local optima found in many nanophotonic applications, while capitalizing on the efficiency of gradient-based optimization.
We demonstrate BONNI's capabilities in the design of a distributed Bragg reflector as well as a dual-layer grating coupler through an exhaustive comparison against other optimization algorithms commonly used in literature.
Using BONNI, we were able to design a 10-layer distributed Bragg reflector with only 4.5\% mean spectral error, compared to the previously reported results of 7.8\% error with 16 layers.
Further designs of a broadband waveguide taper and photonic crystal waveguide transition validate the capabilities of BONNI.
\end{abstract}





\section{Introduction}

The design of components for nanophotonic applications is challenging and often involves labor-intensive manual iterations.
Many optimal designs defy human intuition, and the high-dimensional, non-convex parameter spaces involved remain challenging even for automated optimization methods.
Inverse design automates the search for optimal design parameters through an optimization algorithm without human intervention~\cite{molesky2018inverse, noh2023inverse}.
It is categorized into two primary subfields: topology optimization and shape optimization.
Topology optimization modifies material distribution directly across a discretized domain, often involving thousands of variables.
While this approach offers significant geometric flexibility, it poses challenges regarding manufacturing variability \cite{augenstein2020inverse, raza2024fabrication} and the integration of strict fabrication constraints~\cite{piggott2017fabrication, chen2020design, schubert2022inverse, schubert2025quantized, dai2025shaping}.
Consequently, we focus on shape optimization, where the geometry is defined through explicit parameterization with a small number of design variables.
This problem is formulated as finding the optimal design vector  $x^* \in \arg\max_{x\in \mathcal{X}} f(x)$, where $\mathcal{X} \subsetneq \mathbb{R}^d$ represents the feasible design space.

Optimization strategies for solving this problem generally fall into two categories: global and gradient-based methods.
Global approaches, such as Particle Swarm Optimization (PSO)~\cite{kennedy1995particle} and Genetic Algorithms (GA) \cite{katoch2021review}, rely on stochastic sampling to explore the landscape of $f(x)$.
While robust, these methods suffer from the curse of dimensionality, as the sampling density required to locate solutions grows exponentially with the number of design parameters.
Conversely, gradient-based (or local) methods leverage the gradient $\nabla f$ to guide the search direction.
By utilizing the adjoint method~\cite{pontryagin2018mathematical}, these gradients can be computed at a cost roughly equivalent to one additional simulation.
This makes gradient-based schemes highly efficient for high-dimensional problems.
However, in non-convex landscapes, they are susceptible to convergence at local optima, potentially failing to reach the global optimum $x^*$.
In nanophotonic shape optimization, the objective landscape is usually characterized by a multitude of local optima \cite{marzban2026inverse}.
For example, in Distributed Bragg Reflectors (DBRs)~\cite{wang1974principles, schubert2007distributed}, the interference between reflected and transmitted waves depends cyclically on layer thickness, creating an oscillatory objective function.
Furthermore, the geometric constraints imposed by low-dimensional shape parameterization often restrict the design space, increasing the number of local optima even more.

Various optimization algorithms are employed in nanophotonic inverse design.
This variety stems from a lack of comparative studies and the inherent difficulty of the applications~\cite{schneider2019benchmarking, lee25benchmark, mahlau2025multi}.
A simple local optimization algorithm is gradient descent with the Adam optimizer~\cite{kingma2014adam}, which adaptively estimates the momentum during optimization.
The Adam optimizer is the standard choice for optimizing the parameters of neural networks due to its memory efficiency and capability of working with batched training data.
Another common algorithm is L-BFGS, which is a quasi-Newton algorithm estimating the Hessian matrix using limited memory during optimization~\cite{liu1989limited}.
Alternatively, the Method of Moving Asymptotes (MMA)~\cite{mma} generates a sequence of convex separable subproblems using rational function approximations, where the curvature is controlled by adjusting the position of vertical asymptotes at each iteration.
This approach is popular in inverse design due to the ability of incorporating non-linear constraints as well as quick convergence.

Complementing these local methods, a variety of global optimization approaches are utilized in nanophotonics.
PSO~\cite{kennedy1995particle} is a population-based algorithm, in which a swarm of candidate solutions iteratively explores the search space by adjusting their velocities according to their own best-known positions as well as the global optimum.
Since PSO does not include gradient information, it can be employed in applications where the gradient is difficult or impossible to obtain.
However, as a consequence, convergence is also typically slower than that of gradient-based algorithms, especially in high-dimensional applications.
Another gradient-free global optimization approach is the Covariance Matrix Adaptation Evolution Strategy (CMA-ES)~\cite{hansen1996adapting}.
It is a stochastic algorithm sampling candidate solutions from a multivariate normal distribution.
During optimization, it adaptively updates the covariance matrix to learn the correlations and correct scaling of the objective landscape.
Bayesian optimization takes a more principled approach by learning a surrogate model of $f$ using previous samples, which is used to determine the optimal next sampling point.
Bayesian optimization is a well-studied algorithm with provable convergence guarantees, where its application to high-dimensional applications is an emerging field \cite{hvarfner2024vanilla, papenmeier2025understanding, doumont2025we}.
For example, TuRBO incorporates local probabilistic models into the optimization loop by maintaining a trust region \cite{eriksson2019scalable}.

However, due to the need to sample the entire design space to build an effective surrogate model, gradient-free Bayesian optimization does not scale well to applications with more than approximately 10 design parameters, which includes the majority of nanophotonic design tasks.
One approach to alleviate the burden of many high-fidelity simulations, which are costly even using GPU-acceleration \cite{tidy3d, mahlau2025flexible}, is using multi-fidelity optimization strategies \cite{lu2022multifidelity}.
Even though this reduces simulation cost, balancing the simulation speed with accurate results can be difficult.
Therefore, we focus on high-fidelity simulations and propose BONNI, a global optimization approach capable of efficiently incorporating gradient information into the optimization process.
BONNI combines Bayesian optimization, neural network ensemble surrogate models and Interior Point Optimization (IPOPT)~\cite{ipopt}.
While it is also possible to incorporate gradient information into standard Bayesian optimization, the computational complexity of Gaussian process regression scales cubically in the number of observations.
This issue is further compounded when gradient observations expand the effective dataset size by a factor of $d+1$, limiting its scalability in high-dimensional applications~\cite{garcia2021bayesian, NEURIPS2018_c2f32522}.
In contrast, BONNI leverages the expressivity of neural networks as function approximators for surrogate modeling~\cite{snoek2015scalable, chen2022high, augenstein2023neural}, which scales well to many dimensions.
Internally, BONNI learns the mapping from design parameters to a probability distribution over the figure of merit via the surrogate model.
During optimization, the probability distribution is used to determine the next sampling point by computing the expected improvement over the current best design parameters.
This process of learning a surrogate and determining the next sampling point is repeated until convergence to a global optimum or the computational budget is exhausted.
The full optimization pipeline of BONNI is published open-source\footnote{The BONNI and IPOPT implementation can be accessed at \url{https://github.com/ymahlau/bonni}.} with a Python interface to facilitate its adoption.

We demonstrate the capabilities of BONNI through two inverse design applications.
Our first application addresses the need for distributed Bragg reflectors for indium gallium nitride (InGaN) micro Light Emitting Diodes (µ-LEDs).
Subsequently, we also optimize a dual-layer grating coupler for maximizing transmission efficiency at the interfaces of photonic integrated circuits.
Additionally, we provide further studies on the design of a broadband waveguide taper and photonic crystal waveguide transition in the supplementary material.

\section{Methods}
In optimization applications, one aims to find a maximum $x^* \in \arg\max_{x\in \mathcal{X}} f(x)$ where $f$ is a black box function lacking an analytical closed-form expression.
The domain $\mathcal{X} \subsetneq \mathbb{R}^d$ in our applications is a bounded subset of Cartesian space with $d$ dimensions, defined through a box constraint in every dimension.
For nanophotonic applications, the function $f$ is evaluated through a numerical simulation, which is expensive.
However, despite the high computational cost of simulations, the gradient of $f$ is available through differentiable simulation software~\cite{tidy3d, mahlau2026fdtdx}.
With the adjoint method~\cite{pontryagin2018mathematical}, computing the gradient $\nabla f$ requires approximately the same computational resources as the evaluation of $f$ itself.

\begin{figure}[t]
    \centering
    \includegraphics{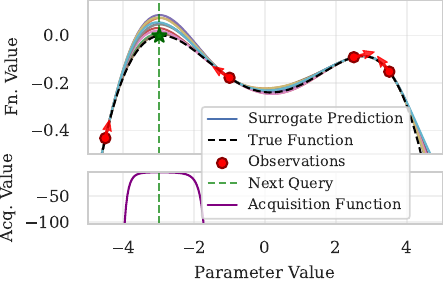}
    \caption{Visualization of the components of BONNI.
    In the upper plot, the surrogate ensemble is trained on four observations (red dots) and gradients (red arrows), given through evaluations of the true function (dashed black line). 
    The individual neural network predictions in the ensemble are visualized through colored lines, forming a confidence measure of the surrogate model.
    In the bottom plot, the expected improvement acquisition value is displayed, which is calculated based on mean and standard deviation of the ensemble predictions.
    The maximum of the acquisition function (green dashed line) is used as the next sampling point for function evaluation (green star).
    After including the newly sampled point in the dataset of sampled points, this process repeats until convergence or computational budget is exhausted.
    }
    \label{fig:bonni}
\end{figure}

\subsection{Bayesian Optimization with Neural Network Ensembles}

Our optimization algorithm BONNI follows similar steps to standard Bayesian optimization with the addition of gradient information.
BONNI trains a surrogate model on $n$ previously observed triplets of input vectors, function values and gradients $\left(x^{(i)}, f(x^{(i)}),  \nabla f(x^{(i)})\right)$ with $1 \leq i \leq n$.
At the beginning of optimization, if no previous observations are available, a fixed number of input values are randomly sampled and evaluated.
The goal of the surrogate model is to approximate the function $f$ and its gradient.
Specifically, the surrogate model learns a mapping from input values $x \in \mathcal{X}$ to a probability distribution over function values $f(x)$.
In gradient-free Bayesian optimization, Gaussian processes are used for this task.
However, while there have been efforts to incorporate gradient information into the Gaussian process surrogates of Bayesian optimization~\cite{garcia2021bayesian}, the computational complexity makes the usage in many applications infeasible.
Instead, we train an ensemble of neural networks~\cite{snoek2015scalable, NIPS2017_9ef2ed4b}.
A neural network ensemble is a set of neural networks that have the same architecture, but different parameters.
Since each network in the ensemble yields a different output, we can quantify the predictive uncertainty by measuring the variance among the model outputs.
Other neural network architectures capable of capturing uncertainty are Bayesian neural networks (BNNs)~\cite{goan2020bayesian, makrygiorgos2025towards}, often implemented through dropout~\cite{gal2016dropout}.
However, we choose ensembles since they exhibit faster and more stable training and prediction qualities compared to BNNs~\cite{gustafsson2020eval}.
Each network in the ensemble consists of $4$ fully connected layers with $256$ hidden units, $8$ normalization groups~\cite{wu2018group} and gelu activation function~\cite{hendrycks2016gaussian} in all but the last layer.
This architecture was chosen to balance model capacity with training speed.
To train the neural networks in the ensemble, their outputs are optimized to fit function values and gradients at the previously sampled input points.
We use $g(x^{(i)}\mid \theta^{(j)})$ with  $1 \leq j \leq m$ to denote the prediction of the neural network with parameters $\theta^{(j)}$ in an ensemble of size $m$ for the sampling point of index $i$.
The loss functions for fitting function values and gradients are 
\begin{align}
    \mathcal{L}_{f} &= \frac{1}{nm} \sum\limits_{i=1}^{n} \sum\limits_{j=1}^{m} \left( f(x^{(i)}) - g(x^{(i)} \mid \theta^{(j)} ) \right)^2, \label{eq:lf} \\
    \mathcal{L}_{\nabla} &= \frac{1}{dnm} \sum\limits_{i=1}^{n} \sum\limits_{j=1}^{m} \left\| \nabla f(x^{(i)}) -  \frac{\partial g(x^{(i)} \mid \theta^{(j)})}{\partial x^{(i)}} \right\|^2_2, \label{eq:ln}
\end{align}

where $n$ is the number of training data points.
The full loss is defined as $\mathcal{L} = \mathcal{L}_{f} + \mathcal{L}_{\nabla}$, which is optimized using the Adam optimizer with decoupled weight decay regularization \cite{kingma2014adam, loshchilov2018decoupled} and a cosine learning rate schedule~\cite{loshchilov2017sgdr}. 
Although a weighting hyperparameter could balance the loss terms, we omit it.
The high capacity of the ensemble allows it to fit both targets with negligible error, rendering weighting unnecessary.
Consequently, all networks in the ensemble learn to closely match function values and gradients at the sampled points.
While this tight fit to observations could in principle reduce diversity between ensemble members near sampled points, we find that the different random initializations~\cite{he2015delving} maintain sufficient predictive diversity in unexplored regions to drive effective exploration.
Within regions far from the sampled data, many different interpolations are plausible such that the variance between the models in the ensemble is high.
In contrast, close to sampled data points, the models are equally constrained, which leads to very similar predictions.
Therefore, the variance between predictions can be used for uncertainty estimation.
In \cref{fig:bonni}, the different model predictions of a trained ensemble are visualized.

\begin{figure}[t]
\centering
\begin{minipage}{0.49\linewidth}
\begin{algorithm}[H]
\caption{BONNI}
\label{alg:bonni}
\begin{algorithmic}[1]
    \Require Domain $\mathcal{X}$, Objective $f$, Gradient $\nabla f$
    
    \State Initialize dataset $\mathcal{D}$ with random samples
    
    \While{budget not exhausted}
        \State \textbf{1. Surrogate Training}
        \State Train NN ensemble $\{g(\cdot \mid \theta^{(j)})\}_{j=1}^m$ on $\mathcal{D}$
        \State Minimize loss \cref{eq:lf} and \cref{eq:ln}
        
        \State \textbf{2. Acquisition Optimization}
        \State $x_{\text{next}} \leftarrow \arg\max_{x \in \mathcal{X}} \text{EI}(x)$ using IPOPT
        
        \State \textbf{3. Evaluation and Update}
        \State $\mathcal{D} \leftarrow \mathcal{D} \cup \{(x_{\text{next}}, f(x_{\text{next}}), \nabla f(x_{\text{next}})\}$
    \EndWhile
    
    \State \Return $x^* \in \arg\max_{(x, \cdot, \cdot)  \in \mathcal{D}} f(x)$
\end{algorithmic}
\end{algorithm}
\end{minipage}
\end{figure}

Subsequently, the utility of a proposed sampling point $x$ is evaluated based on the trained surrogate model.
Given the surrogate's probability distribution over the function values $f(x)$, an acquisition function balances the exploration of regions with high uncertainty and the exploitation of promising regions with low uncertainty.
In Bayesian optimization, a common choice of acquisition function is the Expected Improvement (EI)~\cite{mockus1998application, jones1998efficient}, which we also use for BONNI.
The EI is defined as the expected increase in function value over all previous samples through the new sample.
This is expressed as 
\begin{equation}
    \text{EI}\left(x \mid f(\hat{x})\right) = \mathbb{E}\left[ \max(f(x) - f(\hat{x}), 0 )\right] =\sigma(x) h\left(\frac{\mu(x) - f(\hat{x})}{\sigma(x)}\right),
\end{equation}
where $\hat{x} \in \arg\max_{1 \leq i \leq n} f(x^{(i)})$ is the currently best sampled point.
The function $h$ is defined as $h(z) = \phi(z) + z\Phi(z)$ with $\phi$ and $\Phi$ being the PDF and CDF of a standard normal distribution, respectively.
In our implementation, we use the logarithmic EI~\cite{ament2023unexpected}, which is a numerically stable variant of EI.
We note that the EI acquisition function assumes a Gaussian predictive distribution, whereas the ensemble provides only empirical mean and variance estimates.
For moderately sized ensembles, this approximation may introduce bias.
However, we find it to be effective in practice for guiding the optimization, consistent with prior work on ensemble-based Bayesian optimization~\cite{NIPS2017_9ef2ed4b}.

As a next step, the optimal next sampling point is determined by locating the maximum of EI over the input domain $\mathcal{X}$.
This is an optimization problem of similar complexity to the original problem of optimizing the function $f$.
However, the evaluation of the surrogate $g$ is orders of magnitude faster than the original function $f$, such that this optimization becomes feasible.
We use the IPOPT~\cite{ipopt} algorithm to perform this optimization, which we describe in more detail in the supplementary material.
While we choose IPOPT due to its strong optimization capabilities for this step, in general any optimization algorithm could be used.

After determining the new sampling point $x^{(n+1)} \in \arg\max_{x \in \mathcal{X}} \text{EI}(x \mid f(\hat{x}))$, the expensive function $f(x^{(n+1)})$ and its gradient $\nabla f(x^{(n+1)})$ are evaluated.
Once this new triplet is added to the set of previous observations, the optimization continues by repeating the same process until the time or computational budget is exhausted.
This process is summarized in \cref{alg:bonni}.
Moreover, in \cref{fig:bonni}, the connection between the surrogate model, the acquisition function and the next sampling point is visualized for a single iteration in the BONNI optimization loop.

\subsection{Synthetic Validation}

\begin{figure}
    \centering
    \begin{subfigure}{0.32\linewidth}
        \centering
        \includegraphics{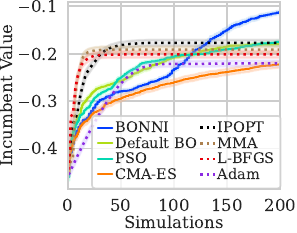}
        \caption{10d}
        \label{fig:bench10}
    \end{subfigure}
    \begin{subfigure}{0.32\linewidth}
        \centering
        \includegraphics{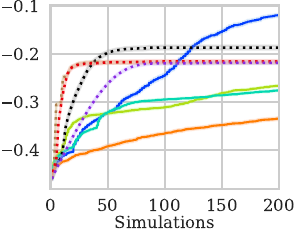}
        \caption{50d}
        \label{fig:bench50}
    \end{subfigure}
    \begin{subfigure}{0.32\linewidth}
        \centering
        \includegraphics{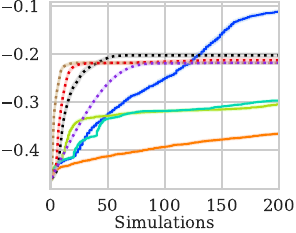}
        \caption{100d}
        \label{fig:bench100}
    \end{subfigure}
    \caption{Comparison of different optimization algorithms on the Rastrigin function in (a) 10, (b) 50 and (c) 100 dimensions.
    For all three experiments, function values are rescaled to the interval $[-1, 0]$.
    The algorithms with dotted lines are local gradient-based algorithms while the solid lines visualize global algorithms.
    We evaluate the incumbent, which is the best value found given a number of simulations.
    All optimizations were performed over 100 random initial configurations.
    The dark lines represent the mean and the shaded regions the standard error over these 100 individual runs.
    }
    \label{fig:benchmark}
\end{figure}

To validate the functionality of BONNI, we test it in a computationally inexpensive benchmark function.
The Rastrigin function with $d$ dimensions is defined as
\begin{equation}
    f_{\text{Rastrigin}} = -10d - \sum_{t=1}^{d}\left(x^2_{t} - 10 \cos\left(2 \pi x_t\right)\right)
\end{equation}
on the domain $[-5.12, 5.12]^d$.
This function contains many local optima with a single global optimum of value zero.
We rescale the function such that the function values are in the range $[-1, 0]$ for better comparison across different numbers of dimensions.

In \cref{fig:benchmark}, the optimization results for the 10-, 50- and 100-dimensional Rastrigin function are displayed.
These dimensionalities represent typical nanophotonic shape optimization applications.
In all three variants, BONNI achieves the best results, validating the advantage of combining global optimization with gradient information.
Notably, the second best results are achieved by IPOPT, indicating its efficacy as a standalone optimizer.
In contrast, gradient-free global optimization fails to converge to competitive solutions for the 50- and 100-dimensional optimization problems.
Only in the 10-dimensional example does Bayesian optimization achieve results comparable to IPOPT within the simulation budget.

\section{Results}

\begin{figure}
    \centering
    \begin{subfigure}{0.45\linewidth}
        \centering
        \includegraphics{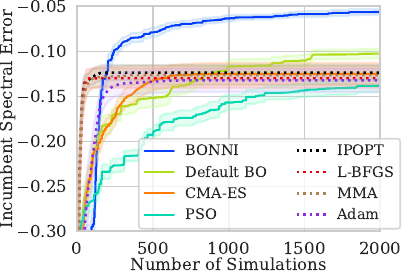}
        \caption{Optimization Results}
        \label{fig:dbg_incumbent}
    \end{subfigure}
    \begin{subfigure}{0.3\linewidth}
        \centering
        \includegraphics{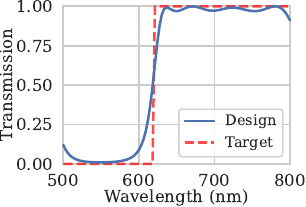}
        \caption{Transmission Spectrum}
        \label{fig:dbg_spectrum}
    \end{subfigure}
    \begin{subfigure}{0.23\linewidth}
        \centering
        \includegraphics{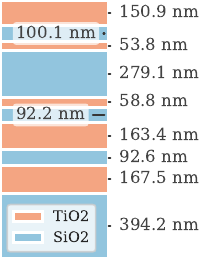}
        \caption{Layer Heights}
        \label{fig:dbg_layers}
    \end{subfigure}
    \caption{Optimization results for the DBR.
    In (a), the optimization results of the different algorithms are visualized.
    The algorithms with dotted lines are local gradient-based algorithms while the solid lines visualize global algorithms.
    We evaluate the incumbent, which is the best value found given a number of simulations.
    All optimizations were performed over 10 random initial configurations.
    The dark lines represent the mean and the shaded regions the standard error over these 10 individual runs.
    In (b), the spectrum of the best design found by BONNI is displayed (blue) and compared to the target spectrum, which is a step function around 620 nm (red).
    In (c), the layer heights for this design of silicon dioxide (blue) and titanium dioxide (red) are visualized.
    }
    \label{fig:dbg}
\end{figure}

We next apply BONNI to real nanophotonic design tasks.
To ensure a rigorous comparison, we use the number of simulations as a metric of computational cost.
Because the adjoint method requires an additional simulation to compute gradients, gradient-based algorithms are limited to half the iteration count of derivative-free methods to reflect the same computational budget.
In all our optimizations, we plot the best result found given a number of simulations in an incumbent plot.
This reflects practical design constraints, where the goal is the best results within a specific computational budget.
For statistically relevant results, we run optimizations multiple times and report mean and standard error.
Due to the computational cost of simulations, the real design tasks are performed over 10 random initial configurations, compared to 100 for the synthetic benchmark.
While this limits statistical power, the consistent ranking of algorithms across seeds and the magnitude of performance differences support the conclusions drawn.
To facilitate reproducibility, we also publish the full code for the simulation setup online.

The performance of an optimization algorithm depends on the choice of hyperparameters.
While hyperparameter optimization can increase performance~\cite{lindauer-dso19a,moosbauer-ieee22a}, it is prohibitively expensive for nanophotonic applications.
To reflect a realistic setting with our benchmarks, we do not perform specific hyperparameter optimization and instead use the default parameters of the respective frameworks.
We report the detailed hyperparameters in the supplementary material.
We note that BONNI incurs additional computational overhead per iteration for surrogate training and acquisition function optimization.
In our experiments, this overhead amounts to approximately 5 seconds per iteration.
For applications involving full wave simulations taking minutes to hours, this overhead is negligible; for inexpensive simulations such as the Transfer Matrix Method (TMM), the overhead becomes a larger fraction of total runtime.

\subsection{Distributed Bragg Reflector}
The fabrication of classical aluminum gallium indium phosphide (AlGaInP) µ-LEDs is difficult to scale to sizes below 20µm due to their strong decrease in efficiency at this scale.
Therefore, InGaN is a preferred technology for red light µ-LEDs, which has a higher efficiency at micrometer sizes \cite{zhuang2021ingan, li24}.
Additionally, InGaN µ-LEDs are less complex to fabricate and have better thermal stability.
However, due to the Quantum-Confined Stark Effect (QCSE), red InGaN µ-LEDs change color under varying current.
Specifically, a blue shift occurs, degrading color accuracy significantly \cite{horng2022study}.
While recent advances in manufacturing techniques for reducing the quantum-mechanical stress of InGaN are able to alleviate the QCSE \cite{cheng2024high}, blue shift remains an issue.

DBRs have been proposed as a possible solution for filtering the undesired shorter wavelengths of the visible spectrum \cite{miao2023modified}.
They are structures of alternating layers with high- and low-refractive index materials.
By adjusting the heights of the different layers, DBRs can control transmission and reflection at different wavelengths \cite{sheppard1995approximate}.
While a larger number of layer pairs allows more fine-grained control over the transmission spectrum, it also increases fabrication cost.
Even though typical DBR designs require more than 10 layers to achieve high efficiencies \cite{ABE2023101183, nano14040349}, we show that it is possible to design a DBR filter with only five layer pairs using BONNI.
In this experiment, we use titanium dioxide and silicon dioxide for a high and low refractive index, respectively.

For simulation, we use the tmm Python library~\cite{byrnes2016multilayer}.
We note that TMM simulations are computationally inexpensive compared to full-wave methods, making this benchmark primarily a test of optimization quality rather than sample efficiency under expensive evaluations.
The refractive indices of titanium dioxide and silicon dioxide are set to $2.5$ and $1.46$ respectively, with a background refractive index of 1.
The target spectrum is an idealized step profile with zero transmission below 620nm and full transmission above this threshold.
We use the average error between the design and the target spectrum as the objective function to minimize, computed at 100 equally spaced wavelength bins between 500 and 800nm.
The objective function is defined as
\begin{align}
    f_{\text{DBR}}(x) = -\frac{1}{|\Lambda|}\sum_{\lambda \in \Lambda} \left| \mathbbm{1}_{\lambda > \hat{\lambda}} - T(x, \lambda) \right|,
\end{align}
where $\mathbbm{1}_c$ is the identity function for condition $c$, $\hat{\lambda}$ is the cutoff wavelength at 620nm and $T(x, \lambda)$ denotes the transmission of the DBR with layer heights $x$ at wavelength $\lambda$.

In \cref{fig:dbg}, the optimization results are visualized.
BONNI achieves the best results, surpassing local optimization algorithms after just a few hundred simulations.
All of the local optimization algorithms quickly stagnate in a local optimum since the design application has a lot of local optima.
Among the local optimization algorithms, IPOPT yields the best results.
Standard gradient-free Bayesian optimization is also able to surpass the local optimizers, but only much later than BONNI.
The best design found by BONNI achieves a mean spectral error of $4.5\%$, corresponding to a peak suppression of -19.9 dB below 620 nm and average transmission of 97\% above 620 nm.
Compared to previously reported mean spectral error of $7.8\%$ with custom-designed 16 layers \cite{miao2023modified}, BONNI is able to achieve better results with fewer layers.

\subsection{Dual-Layer Grating Coupler}

\begin{figure}
    \centering
    \begin{subfigure}{0.47\linewidth}
        \centering
        \raisebox{0.6cm}{\includegraphics[width=\linewidth]{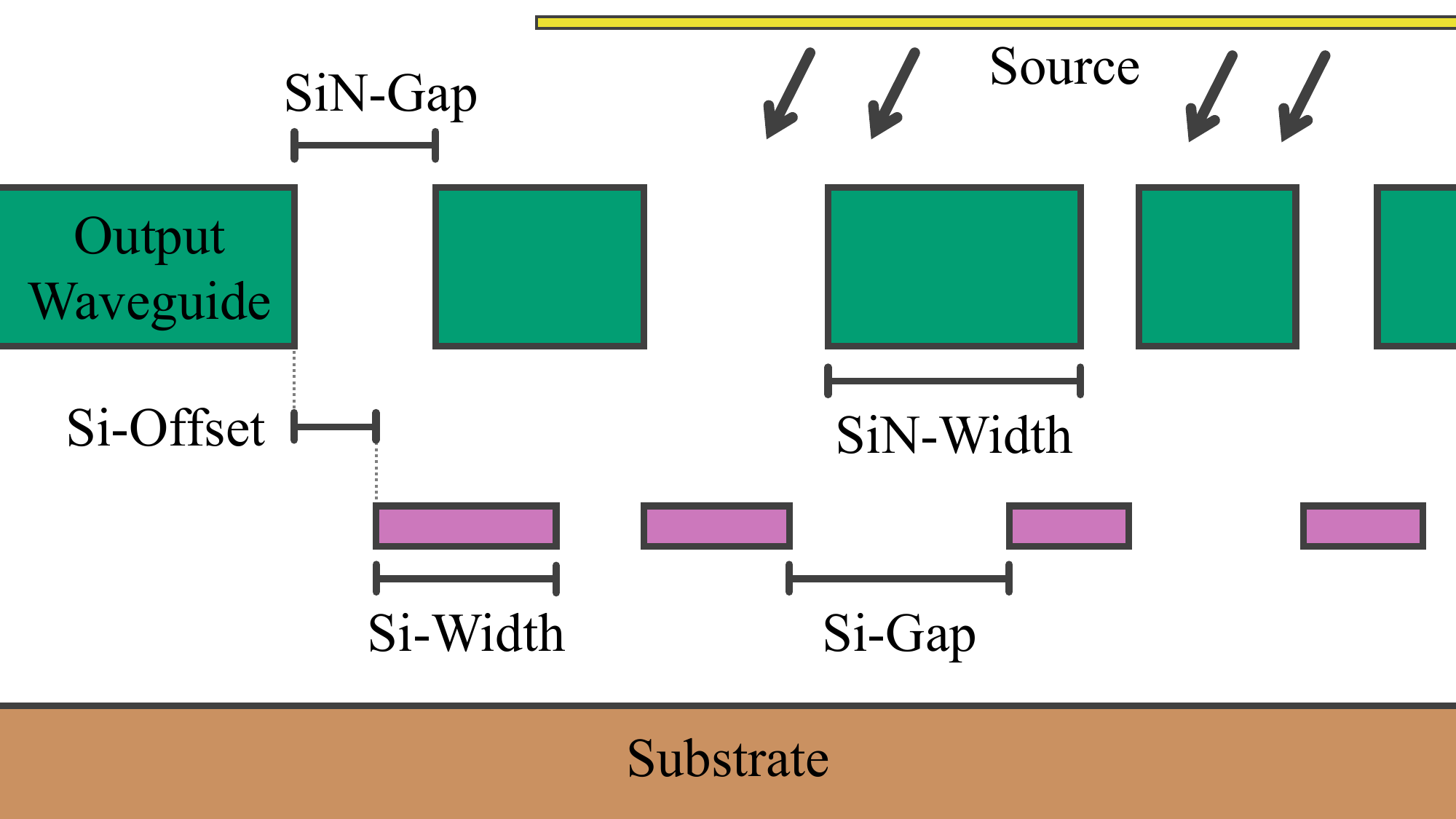}}
        \caption{Parameterization and Simulation Scene}
        \label{fig:gc_setup}
    \end{subfigure}
    \begin{subfigure}{0.5\linewidth}
        \centering
        \includegraphics{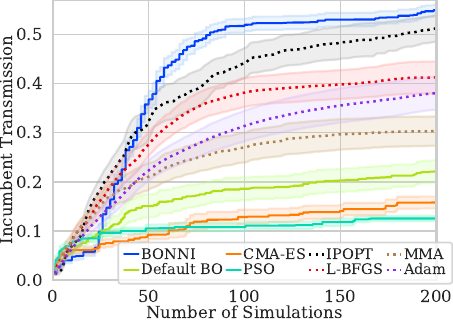}
        \caption{Optimization Results}
        \label{fig:gc_results}
    \end{subfigure}
    \caption{Comparison of different optimization algorithms on the dual-layer grating coupler.
    In (a), the setup of the grating coupler is shown.
    A source (yellow) at the top emits a Gaussian beam onto the two grating layers of silicon Nitride (green) and silicon (pink).
    Transmission is measured at the output waveguide on the right side.
    The two grating layers are placed on top of a silicon substrate (brown).
    In (b), the simulation results are displayed.
    We evaluate the incumbent, which is the best design found given a number of simulations.
    All optimizations were performed on 10 random initial configurations and we report mean (solid line) as well as standard error (shaded area).
    The algorithms with dotted lines are local gradient-based algorithms while the solid lines visualize global algorithms.
    }
    \label{fig:gc}
\end{figure}

A persistent challenge in photonic integrated circuits is the interface between chip and macroscopic optical fibers \cite{photonics12080821, SonHanParkKwonYu18}.
While edge coupling achieves high transmission efficiency, it requires precise alignment and is inherently restricted to the die dicing stage \cite{app10041538, Jang25}.
Therefore, near vertical grating couplers have become a common solution for wafer-scale process control \cite{Marchetti19}.
Since single-layer grating couplers have transmission losses of multiple decibels and enhancements such as reflectors add fabrication costs, dual-layer grating couplers have become an emerging technology \cite{Dai15, Vitali2022}.

Gradient-based design of grating couplers is notoriously difficult as the optimization landscape contains many local optima due to the inherent resonances.
\Cref{fig:gc_setup} displays the parameterization and simulation scene of the grating coupler.
The simulation setup includes a source with wavelength $1.55$µm and a $10$-degree angle to prevent reflections.
Constructed on a silicon substrate with silicon oxide cladding of thickness $2$µm, the lower grating layer consists of silicon with thickness $90$nm.
In contrast, the upper layer consists of silicon nitride of thickness $400$nm, with a distance of $300$nm to the lower layer.
The widths and gaps of the silicon and silicon nitride gratings are the adjustable parameters for optimization.
Additionally, we include parameters for a lateral offset for each of the layers.
With 15 gratings and 15 gaps in both layers, the total number of parameters for this application is 62.
For this benchmark, the figure of merit is the transmission at a wavelength of $1.55$µm.

\begin{figure}[t]
    \centering
    \begin{subfigure}{0.49\linewidth}
        \centering
        \includegraphics{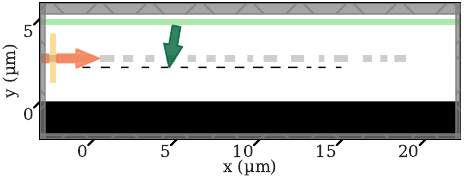}
        \caption{Grating design}
        \label{fig:gc_best_design}
    \end{subfigure}
    \begin{subfigure}{0.49\linewidth}
        \centering
        \includegraphics{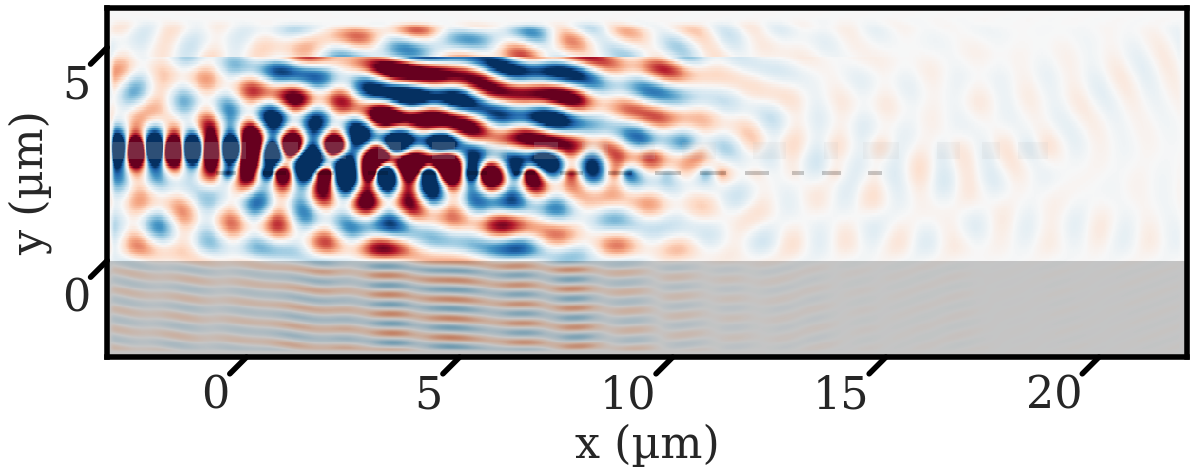}
        \caption{$\text{E}_y$-field distribution}
        \label{fig:gc_best_sim}
    \end{subfigure}
    \caption{Analysis of the best grating coupler design produced by BONNI.
    In (a), the design configuration with gaps and widths of gratings is shown.
    In (b), the simulation result of the design is visualized.
    }
    \label{fig:gc_best}
\end{figure}

In \cref{fig:gc_results}, the results of optimizations are shown.
BONNI achieves the highest transmission efficiency of -2.2 dB with a mean of -2.6 ± 0.3 dB across seeds, followed by IPOPT at -2.9 ± 0.7 dB.
The gradient-based local optimization algorithms, namely L-BFGS, Adam and MMA, only converge to a worse local optimum.
However, they perform better than the gradient-free global optimization algorithms, namely default BO, CMA-ES and PSO, which only find suboptimal solutions with less than 20\% transmission.
In \cref{fig:gc_best}, the best design found by BONNI is visualized.


\section{Discussion}

Our optimization study demonstrates that BONNI consistently yields superior designs compared to the other evaluated optimization algorithms.
It clearly outperforms the alternative global optimization approaches PSO, CMA-ES, and gradient-free Bayesian Optimization across all benchmarks.
The main drawback of BONNI is the training time required for the ensemble surrogate, particularly in the absence of GPU acceleration.
BONNI requires about five seconds to determine the optimal next sampling point when using GPU acceleration.
Without GPU acceleration, the computation time is about five times higher, though specific numbers depend on the hardware setup and algorithm hyperparameters.
While this overhead is negligible when the underlying simulations are time-consuming, it becomes more pronounced during short simulations.
However, even in scenarios involving rapid function evaluations like TMM simulations for the distributed Bragg reflectors, the performance gap between BONNI and competing algorithms is substantial; the superior solution quality justifies the additional computational time between samples.

Remarkably, IPOPT consistently achieved the best performance among all gradient-based algorithms and ranked second overall, trailing only BONNI in nearly all scenarios.
This indicates that IPOPT is a capable optimization approach, despite being rarely utilized in the current nanophotonic literature~\cite{rojas2017short, angeris2021heuristic, swartz2021topology}.
Consequently, we recommend IPOPT over other standard gradient-based approaches as a local optimization algorithm.

The selection between local and global optimization strategies remains problem-dependent.
In high-dimensional parameter spaces with thousands of design variables, local optimization typically outperforms global methods, as it can rapidly converge to a local optimum.
In contrast, global approaches fail to sufficiently explore high-dimensional parameter spaces due to the curse of dimensionality.
Similarly, in smooth optimization landscapes with few local optima, local methods are generally more efficient.
However, most shape optimization applications are characterized by up to 75 design parameters with highly complex, multi-modal landscapes.
In these contexts, global optimization yields superior results than local optimization, provided there is a sufficient sampling budget.

As a practical guideline, we recommend BONNI for design problems featuring a difficult optimization landscape and fewer than approximately 75 parameters, where gradient descent is prone to entrapment in suboptimal local minima.
In very high-dimensional settings, such as topology optimization where training BONNI’s surrogate becomes computationally prohibitive, IPOPT or direct gradient methods remain the pragmatic choice.
Furthermore, in low-dimensional applications with smooth optimization landscapes, IPOPT offers comparable solution quality with lower algorithmic complexity.
Finally, our results emphasize the critical role of gradient information in shape optimization, as shown by the consistently higher performance of gradient-based algorithms over gradient-free alternatives in the 62-dimensional grating coupler benchmark.

From an application perspective, our experiments suggest that substantial performance gains are attainable in many photonics design applications by selecting the appropriate optimization algorithm.
In the design of both distributed Bragg reflectors and grating couplers, BONNI demonstrated the potential to generate superior structures.
Furthermore, the strong performance of IPOPT, which is rarely used in the photonics literature, highlights a lack of rigorous comparative studies in the field of inverse design.

\section{Conclusion}
We addressed the limitations of previous optimization strategies in nanophotonic inverse design by introducing the BONNI algorithm: Bayesian optimization with neural network ensemble surrogates and interior point optimization.
Through two design tasks, we demonstrated that BONNI is the superior choice for applications that have many local optima and are difficult to optimize through gradient descent.
In both the distributed Bragg reflector and the dual-layer grating coupler, BONNI achieved better results than the other algorithms tested.
Using BONNI, we were able to design a distributed Bragg reflector with higher performance using fewer layers than previous results.
Moreover, for constrained computational budgets, IPOPT proves to be a highly efficient standalone optimizer, particularly when the number of local optima is moderate or when simulation cost is low.

Future work will explore the application of BONNI to mixed-variable domains.
A capability of BONNI is its native support for mixed-variable optimization problems containing both differentiable and non-differentiable continuous parameters.
This is achieved by masking the gradient loss for parameters where gradients are unavailable, while still leveraging gradient information for the remaining parameters.
To the best of our knowledge, BONNI is unique among Bayesian optimization methods in offering this capability.
While both applications presented here are fully differentiable, this mixed-variable support holds significant potential for applications such as radio-frequency antenna design, where material choices or discrete structural features coexist with continuously tunable geometric parameters.
Demonstrating this capability on mixed-variable photonic design problems is a primary direction for future work.

\section*{Acknowledgements}

The authors thank Lukas Berg and Jari Luca Gl\"u\ss\, for their help reviewing the manuscript.
This work was supported by the Federal Ministry of Education and Research (BMBF), Germany, under the AI service center KISSKI (grant no. 01IS22093C), the Deutsche Forschungsgemeinschaft (DFG) under Germany’s Excellence Strategy within the Clusters of Excellence PhoenixD (EXC2122) and Quantum Frontiers-2 (EXC2123) , the European Union  under grant agreement no. 101136006 – XTREME.
Additionally, this work was funded by the Deutsche Forschungsgemeinschaft (DFG, German Research Foundation) – 517733257.

\section*{Supporting information}

The following files are available free of charge.
\begin{itemize}
  \item Supplementary Information: Further experiments, analysis and a detailed list of the hyperparameters used in our optimizations.
\end{itemize}

\printbibliography

\appendix
\newpage

\section*{BONNI Hyperparameters}

The proposed BONNI algorithm utilizes a neural network ensemble surrogate and  interior point optimization for the acquisition function.
The specific hyperparameters used for the ensemble training and the internal optimization loop are detailed in Table \ref{tab:bonni_params}.
For the distributed Bragg reflector, we performed 1000 iterations of BONNI starting with 50 random samples.
For the dual-level grating coupler, 100 iterations with 10 random samples were used.
BONNI trains an ensemble of multi-layer perceptrons, where the input is transformed by a normalized embedding mapping the parameter range as $\text{embedding}(x) = (z, cos(z), sin(z))$ where $z = (x - x_{\text{min}}) / (x_{\text{max}} - x_{\text{min}})$.
Additionally, the target values are normalized to a mean of zero and a standard deviation of one.
For acquisition function optimization, we run IPOPT multiple times from different starting points.
The starting points are selected by evaluating the acquisition function on a number of random points and choosing the best as a starting point.

\begin{table}[h]
    \centering
    \renewcommand{\arraystretch}{1.2}
    \begin{tabular}{@{}llp{6cm}@{}}
        \toprule
        \textbf{Component} & \textbf{Parameter} & \textbf{Value} \\
        \midrule
        \textbf{Ensemble} & Ensemble Size ($m$) & 100 \\
                          & Network Architecture & 4-Layer MLP, 256 hidden channels \\
                          & Activation Function & GeLU~\cite{hendrycks2016gaussian} \\
                          & Group Normalization~\cite{wu2018group} & 8 Groups \\
                          & Initialization & He Normal~\cite{he2015delving} \\
                          & Embedding Channels & 3 \\
        \midrule
        \textbf{Training} & Optimizer & AdamW~\cite{kingma2014adam, loshchilov2018decoupled} \\
                          & Learning Rate & peak 1e-3 to 1e-9 \\
                          & Scheduler & Cosine Annealing~\cite{loshchilov2017sgdr} \\
                          & Epochs per Iteration & 1000 \\
                          & Batch Size & All sampled points so far \\
        \midrule
        \textbf{Acquisition} & Inner Opt. Max Iter & 200 \\
                             & Num parallel runs & 10 \\
                             & Random start samples & 100 \\
        \bottomrule
    \end{tabular}
    \caption{Hyperparameters for the BONNI Algorithm.}
    \label{tab:bonni_params}
\end{table}

\section*{Baseline Algorithm Configurations}
To ensure fair comparison, we utilized standard open-source implementations for baseline algorithms. Unless otherwise noted, default parameters provided by the respective frameworks were used.
The default Bayesian optimization, IPOPT, L-BFGS and MMA algorithms do not have specific hyperparameters.
For the Adam optimizer, we used a fixed learning rate of 0.01 without weight regularization.
For the CMA-ES algorithm, we used $\sigma_0=0.2$ as a hyperparameter in the PyCMA framework~\cite{nikolaus_hansen_2025_17765087}.
For PSO, we used $c_1= 0.5$, $c_2=0.3$ and $w=0.9$ with 20 particles in the PySwarms framework~\cite{pyswarmsJOSS2018}.

\section*{Experiment Details}
The distributed Bragg reflector consists of 5 layer pairs of titanium dioxide and silicon dioxide.
For titanium dioxide, we used a refractive index of $2.5$ and for silicon dioxide $1.46$.
The titanium dioxide layers heights were restricted between $26.6$ and $240$ nm.
The silicon dioxide layer heights were restricted between $45.5$ and $410$ nm.
For the dual-layer grating coupler, we restrict both the silicon gap and the grating width to the range between $0.1$ and 1 µm.
For the silicon nitride layer, we restrict the gap widths between $0.3$ and 1 µm, while the widths are restricted between $0.2$ and 1 µm.

\section*{Interior Point Optimization}
We use IPOPT~\cite{ipopt} for optimizing the acquisition function to determine the optimal next sampling point.
IPOPT is well suited for this inner optimization because the expected improvement surface is smooth and differentiable through the neural network ensemble, and the design domain $\mathcal{X}$ is defined by simple box constraints, a setting where IPOPT's barrier method is particularly efficient.
IPOPT is an algorithm that solves general nonlinear optimization problems of the form 

\begin{align}
    \max_{x \in \mathbb{R}^d} \quad f(x) \quad \text{s.t.} \quad x \ge 0.
    \label{eq:ipopt_base}
\end{align}
Any form of box constraints defining $\mathcal{X}$ can be converted into the form of \cref{eq:ipopt_base} using slack variables.
Moreover, it is also possible to incorporate nonlinear constraints of the form $c(x) = 0$, but these are not required for our applications.
Instead of solving the original optimization problem, IPOPT solves a series of simpler barrier problems given by
\begin{align}
    \max_{x \in \mathbb{R}^d} \quad & \varphi_{\mu}(x) := f(x) + \mu \sum_{t=1}^{d} \ln(x_t),
    \label{eq:barrier}
\end{align}
where $x_t$ denotes the entry of index $t$ in the $d$-dimensional vector $x$.
The parameter $\mu$ represents the barrier strength and is annealed to zero during optimization using an adaptive scheduling based on the Karush-Kuhn-Tucker conditions~\cite{MR47303, karush1939minima} for \cref{eq:ipopt_base}.
To solve the barrier problems, IPOPT employs a variant of damped Newton's method utilizing a line-search approach~\cite{NocedalWright2006}.
Specifically, to determine the search direction, the primal-dual formulation of \cref{eq:barrier} is linearized and solved iteratively~\cite{fletcher2002nonlinear}.
This approach has been proven to globally converge under specific assumptions~\cite{Wachter2005}.
To accelerate convergence, IPOPT also employs several enhancements to the scheme described above.
For details, we refer to the original paper~\cite{ipopt}.
Although we employ IPOPT primarily for acquisition function optimization, it has the potential to be a potent standalone optimizer, despite its scarcity in nanophotonic generative design literature.

\section*{Additional Experiments}
We perform additional experiments to test the limits of BONNI in more applications.
Both the following broadband waveguide taper and photonic crystal waveguide taper favor local gradient-based optimization.
The broadband waveguide taper application has a simple optimization landscape with few local optima.
The photonic crystal waveguide taper has 90 dimensions, which makes global optimization difficult.
Moreover, we analyze the effect of the neural network architecture on the optimization results in the distributed Bragg reflector application.

\subsection*{Broadband Waveguide Taper}
A simple application with few parameters is the design of a broadband waveguide taper.
Its purpose is achieving optimal coupling between the modes of a small and large waveguide of widths $450$nm and $4.5$µm.
Parameterized through a list of 30 anchor points representing the distance from the center line, the design shape is derived by fitting a cubic spline through them.
The waveguide consists of silicon surrounded by silicon dioxide, with a waveguide height of $220$nm.
To optimize for worst-case broadband transmission, the figure of merit $f$ is defined as 

\begin{align}
    f(x) = \min_{\lambda} \, T(x, \lambda),
\end{align}
where $\lambda$ is the wavelength in the range between $1$ and $1.5$µm and $T$ is the transmission for the given wavelength and design parameters $x$ represented as a vector of 30 anchor points.

\begin{figure}
    \centering
    \begin{subfigure}{0.49\linewidth}
        \centering
        \includegraphics{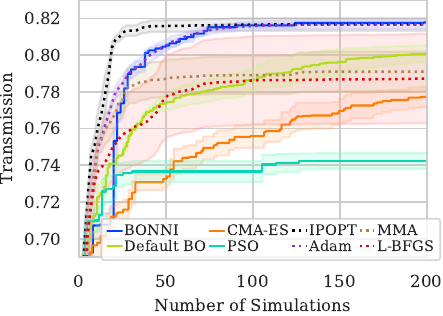}
        \caption{Optimization results}
        \label{fig:taper_results}
    \end{subfigure}
    \begin{subfigure}{0.45\linewidth}
        \centering
        \includegraphics[width=\linewidth]{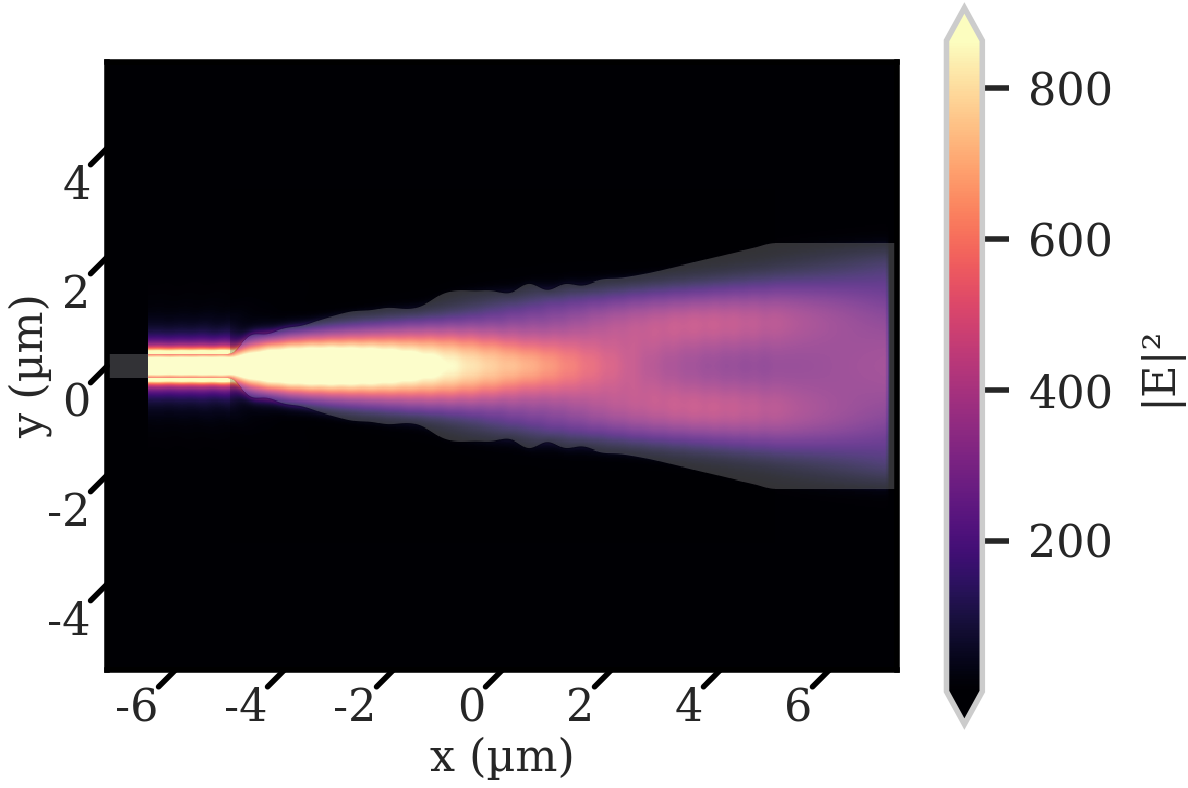}
        \caption{Energy distribution in optimized design}
        \label{fig:taper_sim}
    \end{subfigure}
    \caption{Optimization results for the broadband waveguide taper
    In (a), the performance of different optimization algorithms for this application is visualized. 
    All optimizations were performed on 5 random initial configurations and we report mean (bold line) as well as standard error (shaded area).
    The algorithms with dotted lines are local gradient-based algorithms while the solid lines visualize global algorithms.
    In (b), the best design found by BONNI with its energy distribution is shown.
    }
    \label{fig:taper}
\end{figure}

\Cref{fig:taper_results} presents the quantitative results of the different optimization algorithms.
IPOPT converges to a good solution very quickly, while L-BFGS and MMA fail to find a good solution.
Given enough time, BONNI and default gradient descent eventually match the performance of IPOPT.
The convergence speed of gradient descent could be improved by tuning the learning rate, though this may also lead to divergence or premature convergence to a local optimum.
In contrast, IPOPT and BONNI do not rely on learning rate tuning.
After the full simulation budget, BONNI has the best performance on average, though differences are small.
Similar to the benchmark in the previous section, both PSO and CMA-ES do not find a good solution.
A comparison between BONNI and default BO shows the advantage of gradient information in the optimization process.
The default BO converges more slowly to an optimal solution compared to BONNI and does not reach it within the simulation budget.
\Cref{fig:taper_sim} visualizes the best design found by BONNI.

\subsection*{Photonic Crystal Waveguide Transition}

\begin{figure}
    \centering
    \begin{subfigure}{0.4\linewidth}
        \centering
        \includegraphics{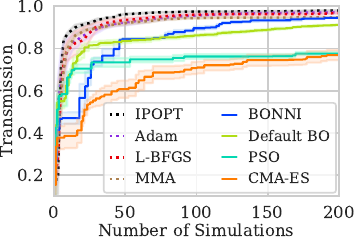}
        \caption{Optimization results}
        \label{fig:pcr_results}
    \end{subfigure}
    \begin{subfigure}{0.55\linewidth}
        \centering
        \includegraphics[width=\linewidth]{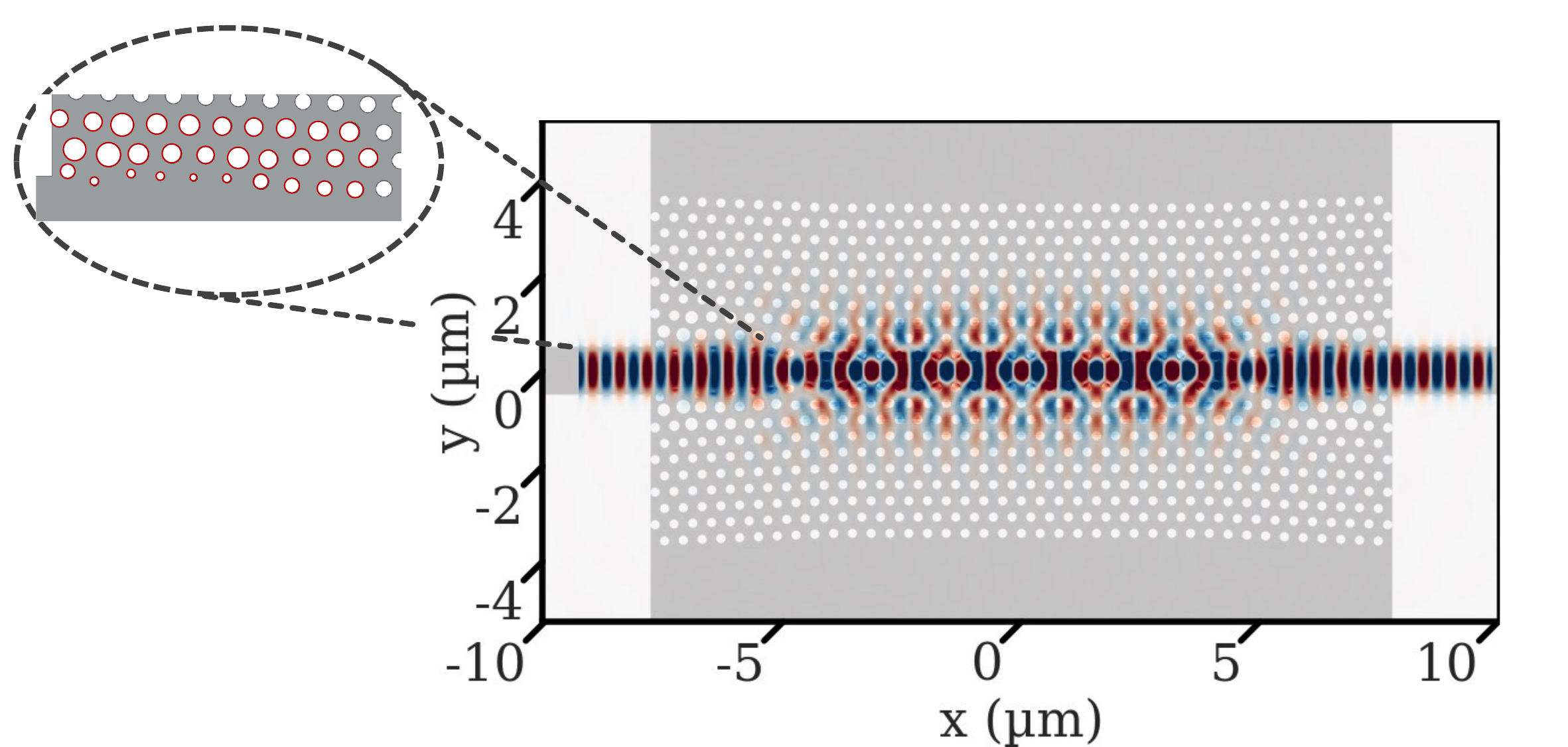}
        \caption{Photonic crystal design}
        \label{fig:pcr_best_sim}
    \end{subfigure}
    \caption{Optimization results for the photonic crystal waveguide transition.
    In (a), the simulation results are displayed. All optimizations were performed on 5 random initial configurations and we report mean (solid line) as well as standard error (shaded area).
    The algorithms with dotted lines are local gradient-based algorithms while the solid lines visualize global algorithms.
    In (b), the optimal design found by IPOPT is visualized with its $\text{E}_y$-distribution in simulation.
    The position and radii of the first ten holes in the first three rows (red circles) can be optimized.
    }
    \label{fig:pcr}
\end{figure}

Photonic crystal waveguides are structures that control the flow of light using a periodic arrangement of materials with different refractive indices.
By introducing a line defect into this crystal lattice, they confine light within a specific path, preventing it from escaping into the surrounding lattice due to the photonic bandgap effect.
Their main use cases are slow light~\cite{baba2008slow} and high quality factor for light modulation~\cite{panuski2022full}.

We optimize the transition between a silicon strip waveguide and a photonic crystal waveguide.
The photonic crystal structure has a lattice constant of 394 nm and a hole radius of 96 nm, while the strip waveguide has a width of 1.002 µm and a height of 210 nm.
This is similar to previously reported setups~\cite{hirotani2021si, shiratori2021particle}.
In the baseline configuration, the first ten holes in the transition area are tapered, which increases transmission~\cite{terada2017optimized}.
During optimization, the position and radius of the first ten holes in the first three rows are treated as design variables for optimization.
With 30 holes being optimized, this results in 90 degrees of freedom for the optimization.
Specifically, the positions of the holes are constrained to a maximum shift of 96 nm in each of the x and y directions.
For the hole size, the radius can be optimized between 40 and 150 nm.
The figure of merit for this benchmark is the transmission of the TE-mode at wavelength $1.55$µm.

\Cref{fig:pcr_results} displays the results of the optimizations.
In this benchmark, local optimization algorithms perform best due to the high dimensionality of the design space, which makes global optimization difficult.
From the local optimization algorithms, IPOPT demonstrates the best results, followed by Adam, L-BFGS and MMA.
BONNI yields the highest transmission among the global optimization algorithms, followed by standard BO.
Following the same trend from the previous sections, CMA-ES and PSO have inferior performance compared to the other algorithms.
In \cref{fig:pcr_best_sim}, the best design found by IPOPT for the photonic crystal waveguide transition is shown.

\end{document}